\documentclass[twocolumn]{aastex6}

\usepackage{breakurl}
\shorttitle{Distinguishing Between BBH Formation Channels with LISA}
\shortauthors{Breivik, Rodriguez, Larson, Kalogera, Rasio}

\begin{document}


\title{Distinguishing Between Formation Channels for Binary Black Holes with LISA}



\author{Katelyn Breivik\altaffilmark{1}, Carl L.~Rodriguez\altaffilmark{1}, Shane L.~Larson\altaffilmark{1,2},  Vassiliki Kalogera\altaffilmark{1}, Frederic A.~Rasio\altaffilmark{1}}


\altaffiltext{1}{Center for Interdisciplinary Exploration and Research in Astrophysics (CIERA)
and Dept.~of Physics and Astronomy, Northwestern University
2145 Sheridan Rd, Evanston, IL 60208, USA}

\altaffiltext{2}{The Adler Planetarium, 1300 S Lakeshore Dr, Chicago, IL 60605}

\begin{abstract}
The recent detections of GW150914 and GW151226 imply an abundance of stellar-mass binary-black-hole mergers in the local universe.  While ground-based gravitational-wave detectors are limited to observing the final moments before a binary merges, space-based detectors, such as the Laser Interferometer Space Antenna (LISA), can observe binaries at lower orbital frequencies where such systems may still encode information about their formation histories.  In particular, the orbital eccentricity and mass of binary black holes in the LISA frequency band can be used together to discriminate between binaries formed in isolation in galactic fields and those formed in dense stellar environments such as globular clusters.  In this letter, we explore the orbital eccentricity and mass of binary-black-hole populations as they evolve through the LISA frequency band.  Overall we find that there are two distinct populations discernible by LISA.  We show that up to $\sim90\%$ of binaries formed either dynamically or in isolation have eccentricities measurable by LISA.  Finally, we note how measured eccentricities of low-mass binary black holes evolved in isolation could provide detailed constraints on the physics of black-hole natal kicks and common-envelope evolution.
\end{abstract}




\section{Introduction} 
\label{sec:intro}

The era of gravitational wave (GW) astrophysics began with the discovery of the binary black hole (BBH) merger, GW150914, by Advanced LIGO (aLIGO)  \citep{Abbott2016b}.  The subsequent detection of BBH merger, GW151226, with smaller progenitor masses suggests diversity in the potential formation channels of the BBHs \citep{Abbott2016c}.  Parameter estimation analyses of GW150914 were done assuming a circular orbit, but eccentricities of $e\lesssim0.1$ would not have been measurable for the event \citep{Abbott2016a}.  Multiple scenarios have been proposed to produce GW150914-like BBHs \citep{Rodriguez2016a,Belczynski2016,Marchant2016,Mandel2016,Bird2016}. Despite the radically different environments of the proposed formation channels, aLIGO is not able to discriminate between them without a full population of measured spins \citep{Vitale2015}, mass ratios, redshift distributions or eccentricities.

BBHs formed dynamically in dense stellar environments like globular clusters (GCs) generally exhibit higher eccentricities than BBHs evolved in isolated environments in galactic fields. Binary evolutionary processes like tidal circularization and mass transfer affect the orbital parameters of BBHs formed in galactic fields \cite[e.g.][and references therein]{Postnov2014}. Once the BBHs are formed, GW emission shrinks and circularizes the initial orbit over long timescales, leading to low eccentricities in both the LISA and aLIGO bands. In GCs, BBHs are formed dynamically and subsequently ejected from the cluster with eccentricities distributed thermally after initially sinking to the cluster core \citep{Morscher2013}. The distribution of BBH separations is a function GC mass and leads to BBHs that evolve through GW emission to the LISA frequency band with non-negligible eccentricities \citep{Rodriguez2016}. 

It has recently been suggested that hundreds of BBHs similar to the progenitor of GW150914 will be observable by a space-based gravitational-wave detector like LISA and subsequently evolve into the aLIGO band within 10 years \citep{Sesana2016}. It has also been shown that BBHs detected by LISA may also have measurable eccentricity if $e\gtrsim0.001$ \citep{Seto2016,Nishizawa2016}. Thus, LISA may be able to discriminate between BBH formation channels using eccentricity. 

In this letter we show that BBHs formed in galactic fields and BBHs formed in GCs have different combinations of mass and eccentricity in the LISA frequency band, thus providing a method to distinguish between the two formation channels.  We also show two distinct populations within the BBHs evolved in isolation that are separated by higher and lower eccentricities that will be potentially distinguishable by LISA.\footnote{For an in depth discussion of a LISA-like detector's capabilities of distinguishing between BBH formation channels using eccentricity, see \cite{Nishizawa2016a}.} In $\S$\,\ref{models} we give a brief overview of our simulated BBH populations.  In $\S$\,\ref{LISAecc} we describe the eccentricity distribution of the simulated BBHs, show how each population of BBHs evolves due to GW emission and suggest formation channels potentially distinguishable by LISA. We provide concluding remarks in $\S$\,\ref{Discussion}.

\section{Galactic Field and Globular Cluster Models}
\label{models}
We simulated two populations of BBHs, one evolved in isolation in galactic fields and one formed dynamically in GCs, to explore the evolution of each population from formation frequencies of $f\sim10^{-9}-10^{-7}$ Hz all the way to the upper end of the aLIGO band at $f\sim10^3$ Hz. The same binary evolution physics was used to create both population simulations with stellar dynamics effects turned on and off. 

We used a modified version of the binary stellar evolution code \texttt{BSE} \citep{Hurley2002} with updated stellar evolution prescriptions including BH formation and natal kicks  \citep{Belczynski2002, Fryer2001}, updated metallicity-dependent stellar-wind prescriptions \citep{Belczynski2010, Vink2001, Vink2005}, and additional mechanisms to account for fallback in neutrino-driven supernovae\citep{Fryer2012}. For this study, we adopt the fiducial models of \cite{Rodriguez2016} which are based on the most recent stellar evolution prescriptions for galactic fields \citep{Dominik2013}. 

Our population of dynamically-formed binaries is taken from a collection of 48 GC models developed in \cite{Rodriguez2016} with the Cluster Monte Carlo (CMC) code \cite[][and references therein]{Pattabiraman2013}.  CMC employs a statistical approach to stellar dynamics, first developed by \cite{Henon1975}, which enables the modeling of star clusters with significantly greater number of particles than a direct $N$-body simulation, while still employing the necessary physics---single and binary stellar evolution with BSE, 3- and 4-body strong gravitational encounters \citep{Fregeau2007}, and dynamical three-body binary formation \citep{Morscher2013}---to fully characterize the dynamical BBH merger problem. We neglect long-term secular effects \cite[e.g.,][]{Antonini2014, Antonini2015} and relativistic dynamical scattering \cite[e.g.,][]{Samsing2014}, which are expected to contribute to the overall BBH population at the $\sim1\%$ or lower level, according to these studies.

These 48 GC models span a range of initial particle numbers ($N = 2\times10^5,~5\times10^5,~1\times10^6,~\rm{and}~2\times10^6$), initial virial radii ($R_v = 1,2~\rm{pc}$), and stellar metallicities ($Z = 0.25Z_{\odot},~0.05Z_{\odot},~\rm{and}~0.01Z_{\odot}$), with two statistically-independent models generated for each set of initial conditions.  As was done in \cite{Rodriguez2016}, we assume that the GC population of the local universe is comprised of $\sim44\%$ high-metallicity GCs ($0.25Z_{\odot}$), and $\sim56\%$ low-metallicity GCs ($0.05Z_{\odot}$ and $0.01Z_{\odot}$).  We further assign to each GC a tidal radius based on its galactocentric distance, which we assume to be correlated to its stellar metallicity based on observations of the Milky Way and other galaxies \citep{Harris2010}.  We finally assume a log-normal GC mass function, based on recent observations of the GC luminosity function in brightest-cluster galaxies \citep{Harris2014} and a mass-to-light ratio of 2 for old stellar systems \cite[e.g.,][]{Bell2003}.

We draw a sample of BBHs ejected from our 48 GC models by randomly selecting binaries from each model.  The number of binaries selected from a given model is determined by its weight, which we assign by dividing the GC mass function into bins, the midpoints of which are set by the average mass of GC models with the same initial particle number.  The integral of the GC mass function over that bin then determines the weight assigned to that model.  In other words, GC models with larger masses ($3\times10^5M_{\odot}$ to $6\times10^5M_{\odot}$, corresponding to models with initial particle numbers of $1\times10^6$ and $2\times10^6$) contribute more binaries than clusters with smaller masses \cite[see][for details]{Rodriguez2016}.  Once we select this population of binaries from all clusters, we generate a five-dimensional Gaussian kernel density estimate (KDE) from the formation masses, separation, eccentricity and ejection times of the BBH population. We sample $1,000$ BBHs from the KDE that evolve to the LISA band at $f=10^{-5}$ Hz in the last Gyr before the present time. We assume that all GCs were formed $12\pm1$ Gyr ago, and we assign for each binary a cluster birth time drawn from a similar Gaussian distribution. The modulated cluster ages allow us to account for the observed spread in ages of GCs\cite[e.g.,][]{Correnti2016}. We require the dynamically-formed BBHs to enter the LISA band in the last Gyr before the present time.

We generate galactic field populations from four sets of $10^{6}$ initial binaries sampled from standard probability distribution functions to assign each binary with an initial metallicity ($Z$), primary mass ($m$), mass ratio ($q$), orbital separation ($a$), and eccentricity ($e$). The population models compare four different metallicities $(Z=1 Z_{\odot},\; 0.25 Z_{\odot},\; 0.05 Z_{\odot},$ and $0.01 Z_{\odot})$,  with the sub-solar metallicities being consistent with the metallicities used in the GC models. 
For the initial primary mass we adopt the stellar IMF $\xi(m) \propto m^{-2.3}, m \geq 1 M_{\odot}$ \citep{Kroupa2001} with a primary mass limit of 150 $M_{\odot}$. We assume a uniform initial mass ratio distribution consistent with current observational constraints \citep{Mazeh1992,Goldberg1994,Kobulnicky2014}. We assume initial orbital separations are distributed uniformly in $\log(a)$ at wide separations ($10R_{\odot}\leq a \leq 5.75\times10^{6}R_{\odot}$) and fall off linearly at small separations as $\zeta(a) \propto (a/a_{0})^{1.2}$, $a < 10R_{\odot}$ \citep{Han1998}. The initial eccentricities are distributed thermally \citep{Heggie1975} as $\eta(e) = 2e$.

We evolve the galactic field population for 13.87 Gyr using the same binary evolution models as the GC population, creating an equivalent population to the GC population but without dynamics. We log the birth parameters of each BBH, including important formation processes like the number of common-envelope episodes and the natal kicks imparted to the binary from the birth of each black hole. As with the dynamically-formed BBHs, we require the low metallicity ($Z < Z_{\odot}$) BBHs to enter the LISA band in the last Gyr before the present time. We retain any solar metallicity BBHs that evolve to the LISA band over the last 10 Gyr. 

Due to the eccentric nature of BBHs formed in GCs, it is useful to consider the frequency of GWs emitted at higher harmonics of a binary's orbital frequency. The frequency of maximum GW power emission from an eccentric binary is estimated as \citep{Wen2003}

\begin{equation}\label{fGW}
f_{GW} = \frac{\sqrt{G(m_1+m_2)}}{\pi}\frac{(1+e)^{1.1954}}{[a(1-e^2)]^{1.5}}.
\end{equation}

If we consider the peak GW frequency of Eq.\,\ref{fGW} for BBHs formed in GCs, we find the GW frequencies of the population are substantially higher than the circular GW frequencies. Since the peak sensitivity of LISA falls near $f\sim10^{-3}-10^{-2}$ Hz, shifts to higher frequency aid in the detectability of these sources.

\section{Eccentricities across the LISA band}
\label{LISAecc}
\subsection{BBH orbital evolution}

Since the dynamically-formed BBHs are ejected from their host GC and the galactic field BBHs evolve in isolation, only GW emission will affect the evolution of each binary. Using the simulated populations from $\S$\,\ref{models}, we evolve each BBH evolved in isolation from its birth frequency to $10^{3}$ Hz and each dynamically-formed BBH from the time of ejection from its host cluster to $10^{3}$ Hz using the quadrupole approximated GW orbital evolution equations \citep{Peters1964}.

\begin{figure*}
\begin{center}
\includegraphics[width=0.60\textwidth]{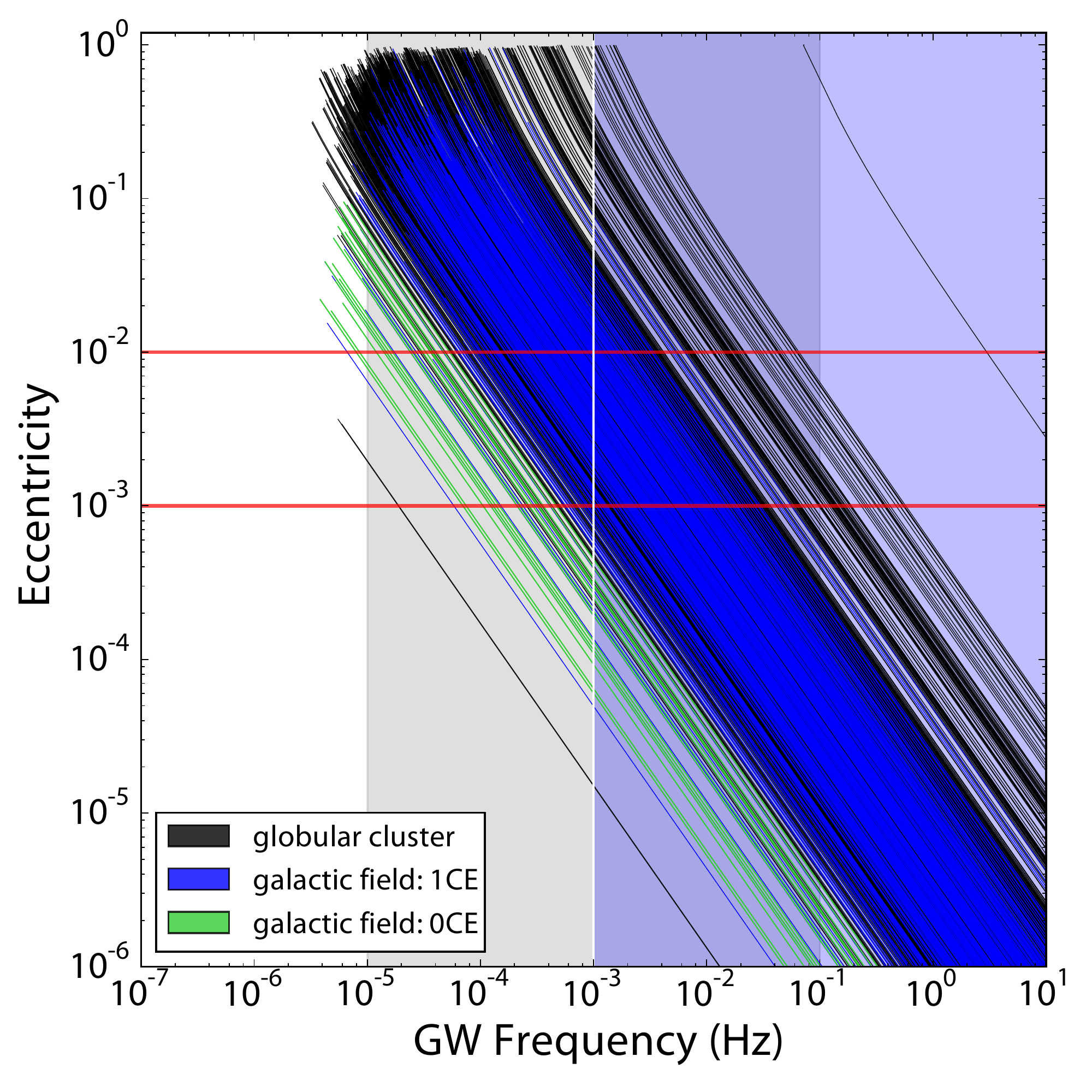}
\end{center}
\caption{Eccentricity evolution tracks as a function of GW frequency for BBHs formed both dynamically in dense stellar environments and in isolation in galactic fields. Black lines denote BBHs ejected from GCs and green and blue lines denote 1CE and 0CE BBHs evolved in galactic fields. The lower horizontal red line denotes the measurable eccentricity ($e\leq0.001$) for 90\% (25\%) of BBHs observed for $T_{obs}=5$ yrs ($2$ yrs).  The upper horizontal red line shows the eccentricity ($e\leq0.01$) that will always be measurable for any observed BBH. The grey band highlights the LISA frequency range and and the blue band highlights the frequency range where BBHs with chirp mass $\mathcal{M}_{c} \gtrsim 6 M_{\odot}$ are expected to have measurable frequency evolution.\label{fig:EccEvolution}}

\end{figure*}

BBHs with frequencies larger than $10^{-3}$ Hz are expected to have measurable frequency evolution due to GW emission known as the GW chirp.  The chirp mass, $\mathcal{M}_{c}=(m_1m_2)^{3/5}/(m_1+m_2)^{1/5}$, and eccentricity (or lack therof) can be measured for every BBH with a measured chirp. Fig.\,\ref{fig:EccEvolution} illustrates the evolution of eccentricity and GW frequency for each binary in the models from each formation channel. The minimal eccentricities measurable by a LISA-like detector are shown in red \citep{Nishizawa2016}. The LISA frequency range is highlighted in grey and the frequency range where BBHs with chirp mass $\mathcal{M}_{c} \gtrsim 6 M_{\odot}$ have measurable chirps is highlighted in light blue.

In addition to dynamically-formed BBHs, we consider two populations of BBHs formed in isolation: those including either a single or no common-envelope episodes (hereafter 1CE and 0CE). BBHs formed both dynamically (black) and in isolation (green, blue) have measurable eccentricities and frequencies above $10^{-3}$ Hz, with dynamically-formed binaries having larger eccentricities than those in galactic fields.  Above $10^{-2}$ Hz, only BBHs formed in GCs fall above the $e=0.01$ line.  This suggests that \textit{any BBH detected above frequency $10^{-2}$ Hz with eccentricity $e\geq0.01$ formed dynamically in a dense stellar environment}.

\subsection{Eccentricity Distributions}
dynamically-formed BBHs are ejected from the cluster with a thermal eccentricity distribution. Once ejected, BBHs evolve only through the emission of GWs which circularize and shrink the binary orbit. This leads to BBHs with eccentricities of $e\gtrsim0.1$ in the low end of the LISA frequency band ($f_{GW}\sim10^{-5}$ Hz) and eccentricities of $e\gtrsim0.001$ in the high end of the LISA frequency band ($f_{GW}\sim10^{-2}$ Hz).  BBHs formed in isolation generally form at lower eccentricities, but with also lower chirp masses than BBHs formed in GCs.  

\begin{figure}
\begin{center}
\includegraphics[width=0.45\textwidth]{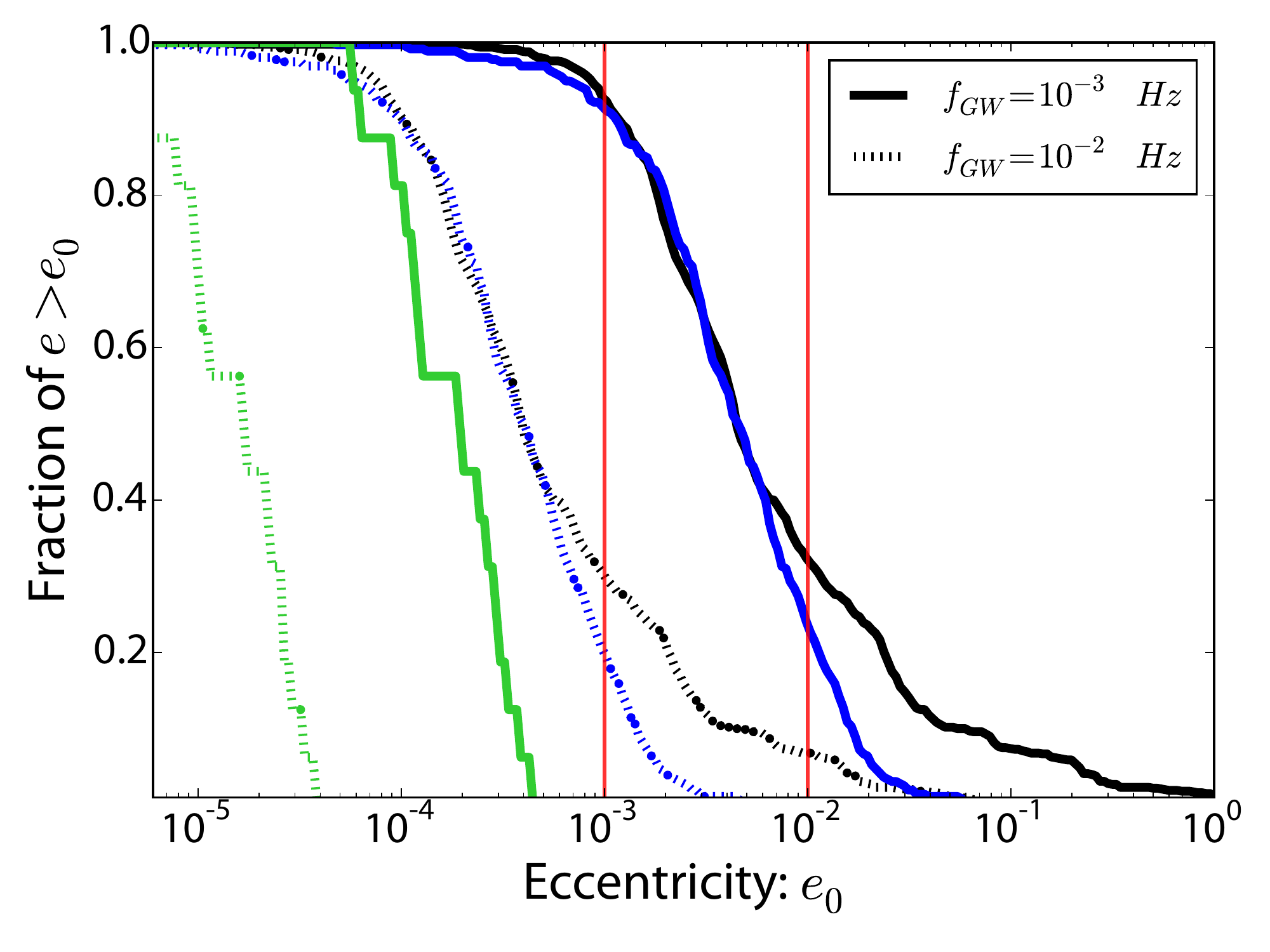}
\end{center}
\caption{Cumulative fraction of dynamically-formed (black), 1CE (blue), and 0CE (green) BBHs with an eccentricity greater than a minimum eccentricity for various points in the orbital evolution of the binaries. The left red line denotes the measurable eccentricity ($e\leq0.001$) for 90\% (25\%) of BBHs observed for $T_{obs}=5$ yrs ($2$ yrs).  The right red line shows the eccentricity ($e\leq0.01$) that will always be measurable for any observed chirping BBH.\label{fig:fracEcc}}
\end{figure}

Fig.\,\ref{fig:fracEcc} shows the cumulative distribution of the eccentricities of the simulated BBHs formed dynamically (black) and in isolation (blue, green) at different points in their orbital evolution. Highly eccentric dynamically-formed binaries are recently ejected from the GC, while the less eccentric binaries have had more time to circularize through GW emission. At $10^{-3}$ Hz, $92\%$ of dynamically-formed BBHs have $e\geq0.001$, while $32\%$ of them have $e\geq0.01$.  At $10^{-2}$ Hz, the eccentricity of the dynamically-formed BBHs has decreased such that $30\%$ have $e\geq0.001$ and $7\%$ have $e\geq0.01$.  For the 1CE BBHs, $91\%$ have $e\geq0.001$ and $23\%$ have $e\geq0.01$ at $10^{-3}$ Hz.  Again, at $10^{-2}$ Hz the eccentricity of the 1CE BBHs has decreased, with $19\%$ having $e\geq0.001$ and $0\%$ with $e\geq0.01$.  There are no 0CE BBHs with measurable eccentricities at GW frequencies above $10^{-3}$ Hz.


\subsection{Chirp Mass and Eccentricity Correlations}
\label{McvsEcc}

\begin{figure}
\begin{center}
\includegraphics[width=0.48\textwidth]{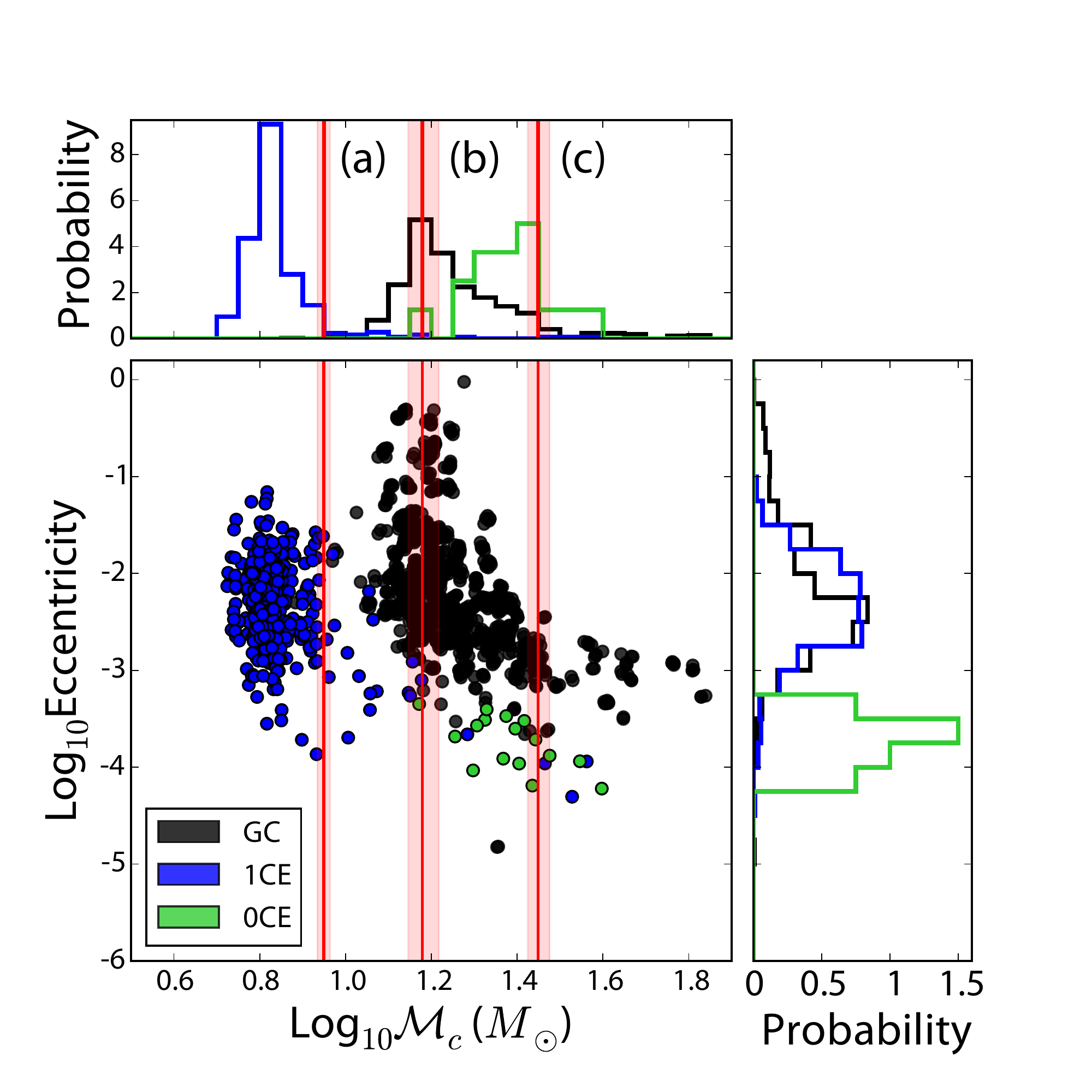}
\end{center}
\caption{Scatter plot and histograms of the chirp mass and eccentricity of the BBHs formed in isolation (green, blue) and dynamically in dense stellar environments (black) at GW frequency $f_{GW}=10^{-3}$ Hz. The red lines denote the chirp mass of the (a) GW151226, (b) LVT151012, and (c) GW150914 progenitors with 90\% confidence limits. The histograms are normalized separately for each population and do not reflect the relative number of sources.\label{fig:mChirpVsEcc}}
\end{figure}

A particularly useful way to separate BBHs formed dynamically or in isolation is through the correlations between the chirp mass and eccentricity of each population.  BBHs with chirp masses $\mathcal{M}_{c} < 10 M_{\odot}$ exclusively form in isolation in galactic fields, though we note that young, high-metallicity clusters (not included in our models) are capable of producing BBHs with lower chirp masses \citep[in prep]{Chatterjee2016}.

If only the chirp mass of a BBH with $\mathcal{M}_{c} > 10 M_{\odot}$ is observed, it is impossible to discern which population it originated in since the chirp masses of the GC, 1CE, and 0CE populations overlap in this region.  However, if the eccentricity is also measured, the three populations can be  resolved. Fig.\,\ref{fig:mChirpVsEcc} shows the eccentricity vs chirp mass plots of each population at $f_{GW}=10^{-3}$ Hz. The shape of the distributions stays constant but the eccentricity decreases as the BBHs evolve to higher frequencies through GW emission.  For each population, the estimation error on the chirp mass is $\Delta\mathcal{M}_c/\mathcal{M}_c\simeq 3(\Delta\dot{f}/\dot{f})/5$ where $\Delta\dot{f}=0.43(SNR/10)^{-1}T_{obs}^{-2}$ \citep{Takahashi2002}.  In all cases, the chirp mass estimation error is smaller than the width of the data points.

The chirp masses of the GW150914 and GW151226 progenitors are plotted in Fig.\,\ref{fig:mChirpVsEcc}.  The GW150914 and LVT151012 progenitors are consistent with both dynamical and isolated formation channels, and the GW151226 progenitor is consistent with the 1CE isolated formation channels.  \textit{If the GW150914, LVT151012, or GW151226 progenitors had been observed by LISA, an eccentricity measurement (or lack thereof) could have aided in identifying their formation histories}.

\subsection{Multiple Field Populations}\label{fieldPops}

\begin{figure*}
\begin{center}
\includegraphics[width=0.95\textwidth]{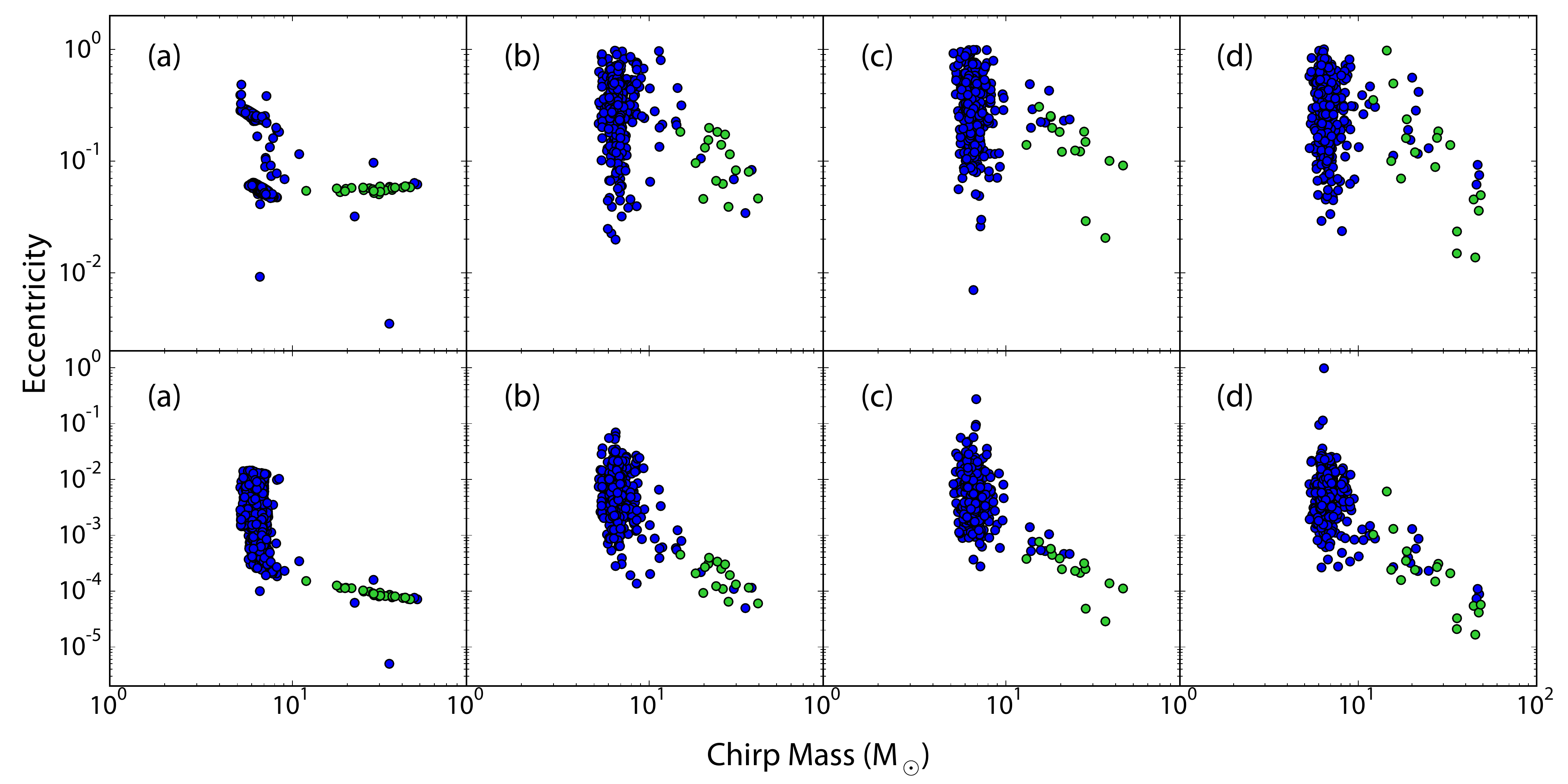}
\end{center}
\caption{Eccentricity vs chirp mass at birth (top row) and $10^{-3}$ Hz (bottom row) of 0CE (green) and 1CE (blue) BBHs formed in isolation in galactic fields. The panels are (a) no natal kicks, (b) fallback modulated kicks, (c) fractional NS kicks weighted by mass, and (d) full NS kicks.  The magnitude of the natal kick increases from right to left.\label{fig:kicksEcc}}
\end{figure*}

We find that the population isolated binaries that evolve into the LISA frequency band is split into higher and lower eccentricity populations. The higher eccentricity population is comprised entirely of BBHs which have undergone a single common-envelope episode while the lower eccentricity population is comprised of a combination of 0CE and 1CE BBHs. The higher-eccentricity 1CE population, while still distinct from the GC BBHs, has potentially measurable eccentricities.

The natal kick imparted to a BH at birth is connected to kicks thought to be imparted to neutron stars (NSs).  NS natal kicks are assumed to follow a Maxwellian distribution with a dispersion of 265 km/s \citep{Hobbs2005}.  The BH natal kick is modified to be dependent on the mass of the pre-collapse stellar core through fallback processes, with higher kicks imparted to lower-mass remnants\citep{Fryer2012}. A binary can be disrupted if the overall energy imparted through the natal kicks is higher than the binding energy of the binary orbit.

In the case of the higher-eccentricity population, the 1CE binaries are driven to small separations through the common-envelope mechanism. Since the masses of the 1CE BBH components are generally low due to low fallback, the natal kick speeds are high enough to produce higher eccentricities, but not so high as to disrupt the BBHs. This suggests that the eccentricity of the 1CE population is dependent on the natal kick physics of the formation of the second BH. 

We simulated three additional populations with varying natal kick prescriptions to explore the effect of natal kicks on the eccentricity of BBHs formed in isolation.  The first prescription sets the BH natal kick equal to zero, thus any orbital changes imparted to the binary are due to momentum conservation from instantaneous mass loss in the BH formation.  We also include two variants of the BH natal kick prescription: one using standard NS kicks for BHs and one that modifies the NS kick by the mass fraction $(M_{NS}/M_{BH})$, where $M_{BH}$ is the mass of the newly formed BH and we assume $M_{NS}=1.4\ M_{\odot}$.

We compare our results for four BH natal kick prescriptions: no natal kicks, fallback modulated kicks, fractional NS kicks weighted by mass, and full NS kicks.  We did not generate new GC simulations for each kick prescription since BBHs formed dynamically in dense stellar environments form with thermally distributed eccentricities, losing the memory of kick effects on the eccentricity distribution.

Fig.\,\ref{fig:kicksEcc} plots the eccentricity vs chirp mass for the 0CE and 1CE formed in isolation for the four natal kick prescriptions mentioned above at their birth orbital frequency and at $10^{-3}$ Hz. Again, in all cases, the chirp mass estimation error bars are smaller than the width of the data points.  The lower mass BBHs in all cases have preferentially higher kicks and thus higher birth eccentricities. The 0CE BBHs generally have larger birth separations and masses which results in low eccentricities by the time the BBH evolves through GW emission to the LISA frequency band.  
 
As expected, BBHs formed with full NS natal kicks retain the largest eccentricities through their GW driven orbital evolution, followed by fractional NS kicks, then fallback modulated kicks and no natal kicks. More massive BBHs evolve to lower eccentricities at a given orbital frequency than the lower mass systems in the same population because of the mass dependence of orbital frequency. This is responsible for diminishing the high eccentricities found at birth in the 0CE population for both the full and fractional NS kicks once they reach the LISA band.

\section{Discussion}\label{Discussion}
We have shown that BBHs formed both dynamically and in isolation may have measurable eccentricity in the LISA band. If BBHs are detected by LISA with eccentricities of $e\gtrsim0.01$ at frequencies above $10^{-2}$ Hz, they will have almost certainly originated from dynamical processes in old, dense stellar environments.  If BBHs with eccentricities of $e\gtrsim0.01$ and chirp masses of $\mathcal{M}_{c} \lesssim 10 M_{\odot}$ are detected by LISA at low frequencies, they likely originated from a common-envelope formation scenario.  

In the future, we plan to extend this study by implementing a full treatment of the formation-redshift distribution of the BBHs observable by LISA originating from both dynamical processes in dense stellar environments and isolated binary evolution in galactic fields.  We also plan to properly account for the detectability of each BBH using eccentricity dependent signal-to-noise ratios \citep[in prep]{Breivik2016a}.   

\acknowledgments
KB and SLL acknowledge support from NASA Grant NNX13AM10G.
CR and FAR acknowledge support from NSF Grant AST-1312945 and from NASA Grant NNX14AP92G.
VK acknowledges support from NSF Grant PHY-1307020/002 and from Northwestern University.
VK and FAR also acknowledge support from NSF Grant PHY-1066293 at the Aspen Center for Physics.

\bibliographystyle{aasjournal}

\end{document}